\documentclass[aps,prb,twocolumn,showpacs]{revtex4}
\usepackage{graphicx,amsmath,textcomp}

\begin{document}
\title{Microscopic model for the novel frustrated Cu(II)-spin
tetrahedron-based Cu$_{4}$Te$_{5}$O$_{12}$X$_{4}$ (X=Cl, Br) systems}

\author{Badiur Rahaman$^{1}$, Harald O. Jeschke$^2$, Roser Valent{{\'\i}}$^2$ and
T. Saha-Dasgupta$^{1}$}

\affiliation{$^1$S.N. Bose National Centre for Basic Sciences, 
JD Block, Sector 3,
 Salt Lake City, Kolkata 700098, India.}

\affiliation{$^2$ Institut f{{\"u}}r Theoretische Physik, Universit{{\"a}}t Frankfurt,
Max-von-Laue-Str. 1, 60438 Frankfurt, Germany}

\date{\today}

\begin{abstract}

{ We present a microscopic study of the electronic and magnetic
properties of the newly synthesized Cu(II)-spin tetrahedron-based
Cu$_4$Te$_{5}$O$_{12}$Cl$_{4}$
compound based on Density Functional calculations and on {\it ab
initio}-derived effective models. In view of these results, we discuss the origin of the
observed differences in behavior between this system and the
structurally similar and much studied 
Cu$_2$Te$_2$O$_5$Cl$_2$.   Since the Br analog of the title compound has
not been synthesized yet,   we derive the crystal structure of
Cu$_4$Te$_{5}$O$_{12}$Br$_4$ by geometry optimization in an {\it ab
initio} molecular dynamics calculation and  investigate the effect of
substituting Cl by Br as well as the
possible magnetic behavior
of this  system  and compare with the recently studied sister compound,
Cu$_2$Te$_2$O$_5$Br$_2$.}

\end{abstract}



\maketitle

\vspace*{0.2cm}


{\it Introduction.-} Frustrated magnetism has gained a lot of
attention in recent years due to the wealth of new exotic behavior
that arises out of this condition such as spin ice and spin liquid
phases~\cite{Moessner06}.  In the search for new materials exhibiting
frustrated magnetism, a few years ago Johnsson {\it et
al.}~\cite{Johnsson00} synthesized for the first time a family of
oxohalogenides Cu$_2$Te$_2$O$_5$X$_2$, X=Br, Cl whose structure was
based on weakly coupled tetrahedra of Cu(II) with
geometrically frustrated antiferromagnetic (AF) interactions.  These
materials have been intensively studied both experimentally and
theoretically~\cite{Lemmens01,Brenig01,Gros03,Valenti03,Jensen03,Kotov04,Zaharko04,Crowe06}. They
show magnetic ordering with incommensurate wave vectors at
temperatures T$_N$ = 18 K (Cl) and 11 K (Br) 
and the observation of a
longitudinal magnon~\cite{Gros03} in Cu$_2$Te$_2$O$_5$Br$_2$ was
interpreted as evidence for the proximity of this system to a quantum
phase transition between antiferromagnet and spin liquid behavior.

 The various intratetrahedral and intertetrahedral couplings and the
 relative strengths of exchange pathways in these compounds have been
 obtained in detail~\cite{Valenti03} by using the electronic structure
 technique of muffin-tin orbital (MTO) based {\it
 N}MTO-downfolding~\cite{nmto}. The results predicted by this study
 have been confirmed by subsequent neutron diffraction
 experiments~\cite{neutron} proving the powerfulness of this {\it ab
 initio} Density Functional Theory (DFT) based technique in
 predicting the underlying microscopic model of a complex material.

\begin{figure}[t]
\includegraphics[width=9cm,keepaspectratio]{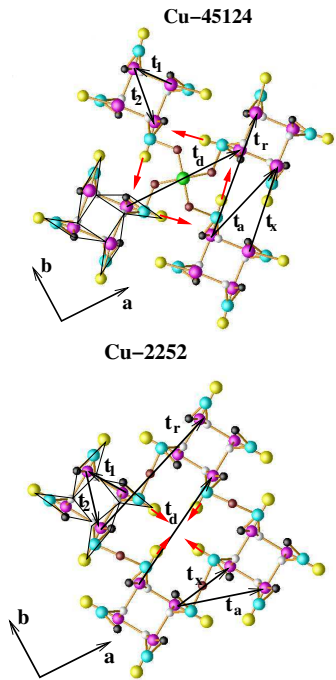}
\caption{(Color online) Crystal structure of  Cu-45124(Cl) (top panel) and
Cu-2252(Cl) (bottom panel)  projected on the $ab$ plane. 
In order to emphasize the similarity between the Cu-2252(Cl) and
Cu-45124(Cl) structures, we use four unit cells of Cu-45124(Cl) but
show only the section that makes this structure analogous to a section
of four Cu-2252(Cl) unit cells showing four connected Cu tetrahedra.
Magenta and yellow atoms stand for Cu and Cl.
Two inequivalent Te atoms, Te(1) and Te(2) in Cu-45124(Cl) are shown
in green and cyan colors (top panel), while the only one inequivalent
Te atom present in Cu-2252(Cl) is shown in cyan (bottom panel).  The
brown, white and black balls denote O(1), O(2) and O(3) respectively
for Cu-45124(Cl) (top panel) and O(3), O(1) and O(2) for Cu-2252(Cl)
\protect\cite{comment_data}.  Note that every four Cu atoms appearing
in a square arrangement, due to the projection, actually form
tetrahedra. We also show the various interaction paths in black arrows
(see the text for discussion). }
\label{struc1}
\end{figure}

By changing the subtle ratio between the various interaction paths in
these materials, for instance by applying pressure or by introducing
chemical modifications~\cite{Lemmens03,Kreitlow05,Wang05}, one can attempt to
drive these systems into quantum criticality.  Following these ideas,
a new oxohalogenide Cu$_4$Te$_{5}$O$_{12}$Cl$_{4}$ has been very
recently synthesized by Takagi {\it et al.}~\cite{peter} which orders
antiferromagnetically at T$_N$ = 13.6 K.  This system is structurally
similar to the previously discussed Cu$_{2}$Te$_{2}$O$_{5}$Cl$_{2}$
compound but presents some markedly different features.  As pointed
out by Takagi {\it et al.}, the primary structural difference between
the new Cu$_{4}$Te$_{5}$O$_{12}$Cl$_{4}$ (which we refer to as
Cu-45124(Cl) following Ref.~\onlinecite{peter}) and
Cu$_{2}$Te$_{2}$O$_{5}$Cl$_{2}$ ( Cu-2252(Cl) ) is the presence of a
TeO$_{4}$ complex in the middle of the Cu-tetrahedral network in the
$ab$ plane (see Fig.\ref{struc1}). This fact led the authors of
Ref.~\onlinecite{peter} to expect an increase in the separation
between the Cu$_4$ tetrahedra and hence an increase in the relative
importance of the intratetrahedral coupling with respect to the
intertetrahedral coupling.

In the following, we study this proposition within the framework of
the {\it N}MTO-downfolding technique. In addition, motivated by the more
anomalous properties of Cu$_{2}$Te$_{2}$O$_{5}$Br$_{2}$ compared to
Cu$_{2}$Te$_{2}$O$_{5}$Cl$_{2}$ as reported in the
literature~\cite{Lemmens01,Brenig01,Gros03,Valenti03,Jensen03,Kotov04,Zaharko04,Crowe06,Lemmens03,Kreitlow05},
we explore the implications of substituting Cl by Br in the new
oxohalogenide Cu-45124(Cl).  Since the Cu-45124(Br) system has not been
synthesized yet, we propose the crystal structure of
Cu$_{4}$Te$_{5}$O$_{12}$Br$_{4}$ by performing a geometry relaxation
in the framework of {\it ab-initio} molecular dynamics and we analyze its
electronic structure by the {\it N}MTO-downfolding technique.  


\begin{figure}[t]
\includegraphics[width=9cm,keepaspectratio]{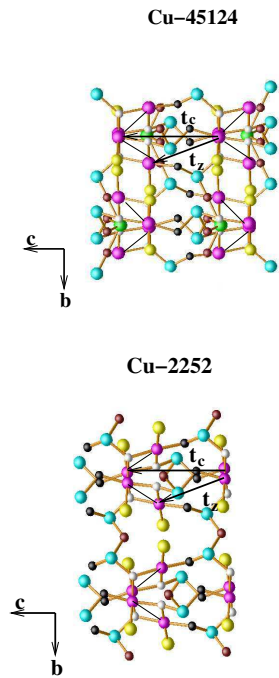}
\caption{(Color online) View of the  crystal structure of the Cu-45124(Cl) (top panel) and
Cu-2252(Cl) (bottom panel) compounds along the  [001] direction. Color scheme is the same as
in Fig.~\ref{struc1}. The thick arrows denote the various interaction paths.}
\label{struc2}
\end{figure}

{\it Structure.-} Both Cu-45124(Cl) and Cu-2252(Cl) compounds
crystallize in a tetragonal structure.  The basic structural unit in
both systems is the [CuO$_{3}$Cl] distorted square (marked in thin
lines in Fig.~\ref{struc1}) with Cu (marked in magenta in
Fig.~\ref{struc1}) at the center. Groups of four such squares share
corners, giving rise to [Cu$_{4}$O$_{8}$Cl$_{4}$] units with Cu ions
in tetrahedral coordination forming a magnetic cluster of four Cu
ions.
Cu-45124(Cl) has two inequivalent Te atoms, Te(1) and Te(2) (marked in
green and cyan in Fig.~\ref{struc1} top panel), while Cu-2252(Cl) has
only one type of Te atoms (marked in cyan in Fig.~\ref{struc1} bottom
panel). The Te(2) atoms in Cu-45124(Cl) sit on an equivalent position
to the Te atoms in Cu-2252(Cl).
Viewing the structures along the [001] direction (see
Fig.~\ref{struc2}), they show a stacking of Cu$_4$ tetrahedra
separated by layers of Te(2)-O(3) (Cu-45124(Cl)) or Te-O(2)
(Cu-2252(Cl)) units. In the case of Cu-45124(Cl), additional
Te(1)-O(1) units appear in the same layer as Cu. The relative
orientation of the Cu$_4$ tetrahedra along the [001] direction is also
different between the two compounds. In the case of Cu-2252(Cl) the
Cu$_4$ tetrahedra show the same orientation, while for
Cu-45124(Cl) they alternate between successive rows. The latter
feature leads to two different space group symmetries, $P\bar{4}$ for
 Cu-2252(Cl) and $P4/n$ for Cu-45124(Cl) with
elongation of the unit cell in the $ab$ plane with lattice parameter
$a = 11.35$~{{\AA}} for Cu-45124(Cl) compared to $a = 7.84$~{{\AA}} for
Cu-2252(Cl). The unit cell dimensions along the $c$ axis remain
comparable with $c= 6.32$~{{\AA}} for Cu-2252(Cl) and $c= 6.33$~{{\AA}} for
Cu-45124(Cl). The change in space group defines Cu-45124(Cl) as
centrosymmetric compared to the non-centrosymmetric Cu-2252(Cl).

A crucial difference between the two compounds apart from the change
in bond lengths, is the relative orientation of the Cu-Cl bonds (Cl
atoms are marked in yellow in Fig.~\ref{struc1}) among different
Cu$_4$ tetrahedra. As has been discussed in
Ref.~\onlinecite{Valenti03}, the Cl atoms play an important role in
mediating the Cu-Cu interaction in the Cu-2252 systems. While for
 Cu-2252(Cl) the Cu-Cl bonds belonging to different Cu$_4$
tetrahedra point towards each other (marked with red arrows in
Fig.~\ref{struc1} bottom panel), in the case of Cu-45124(Cl), due to
the relative shift of the tetrahedra, they are oriented parallel to
each other (see the red arrows in Fig.~\ref{struc1} top panel). This
aspect is found to have important consequences in the context of
hopping interaction pathways as will be discussed later.

\begin{figure}[t]
\includegraphics[width=12cm,keepaspectratio]{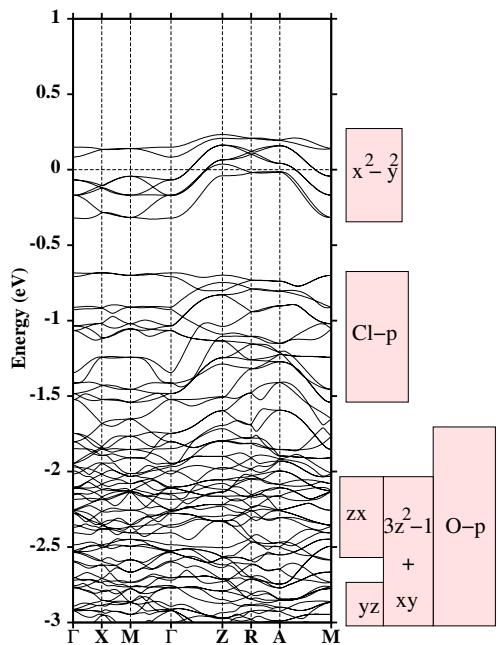}
\caption{LDA band dispersion of Cu-45124(Cl)  plotted along various
symmetry directions with $\Gamma= (0,0,0)$, $X= (\pi,0,0)$, $M=(\pi,\pi,0)$, $Z=
(0,0,\pi)$, $R= (0,\pi,\pi)$ and $A = (\pi,\pi,\pi)$. The dominant orbital
contributions in various energy ranges are shown in small boxes drawn
on the right hand side.}
\label{band}
\end{figure}


{\it Bandstructure.-} Fig.~\ref{band} shows the non spin-polarized 
band dispersion of Cu-45124(Cl) obtained with the linear muffin-tin
orbital (LMTO) basis~\cite{LMTOcode} within the framework of local
density approximation (LDA). The bands are plotted along the
various symmetry directions of the tetragonal Brillouin zone. The
orbital characters indicated in the figure are obtained by choosing
the local coordinate system with the $y$ axis pointing along the
Cu-O(3) bond and the $x$ axis pointing along the Cu-Cl bond. The
square planar symmetry of the ligands surrounding the Cu$^{2+}$ ion
sets the Cu-3$d_{x^2-y^2}$ energy level as the highest energy
level. Consistent with the Cu$^{2+}$ valency, eight bands (there are 8
Cu atoms in the unit cell) dominated by Cu-$d_{x^2-y^2}$ character and
split off from the rest of the bands, span an energy range from $\approx
-0.3$~eV to $0.2$~eV with the zero of energy set at the LDA Fermi
level. The energy bands dominated by other $d$~characters like
$d_{xy}$, $d_{zx}$, $d_{yz}$ and $d_{3z^2-1}$ span the energy range
between $\approx -3$ and $-2$~eV overlapping with the O-$p$ manifold. The
Cl-$p$ dominated bands appear right above and partly overlapping the
O-$p$ dominated bands spanning an energy range of $\approx 1$~eV.  These
Cl-$p$ dominated bands are separated by a gap of $\approx 0.5$~eV from the
Cu-$d_{x^2-y^2}$ dominated bands.
There is only a negligible contribution of Te(1) and Te(2) to the
bands crossing the Fermi energy.  We note that in the low-energy
scale, the LDA calculation leads to eight almost half-filled
bands.  Introduction of correlation
effects within an LDA+U treatment are expected to drive the system
insulating.  In what follows though we will focus on the {\it ab initio}
determination of effective one electron hopping interactions
which are well described within LDA and GGA.

\begin{figure}[t]
\includegraphics[width=9cm,keepaspectratio]{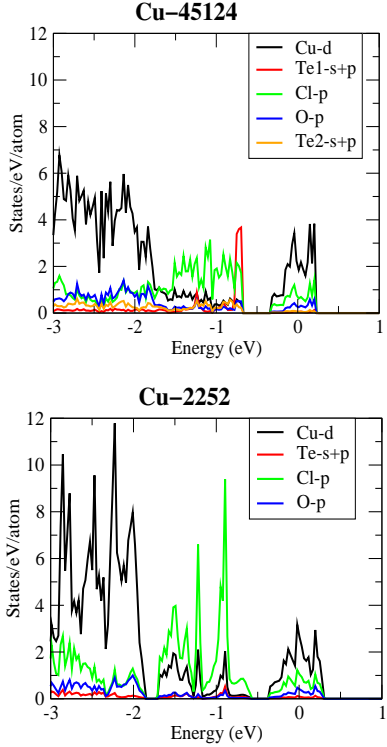}
\caption{(Color online) Comparison of the density of states between the Cu-45124(Cl) (top panel)
and Cu-2252(Cl) (bottom panel)  compounds.}
\label{dos}
\end{figure}

In Fig.~\ref{dos} we show a comparative study of the
various partial LDA density of states (DOS) for  Cu-45124(Cl)
and Cu-2252(Cl).

 While the
basic features of the DOS  remain the same between the
two compounds -indicating that the overall nature of the interactions
will be similar for both systems- there are a few quantitative
differences.  The Cu-$d$ bandwidth at   $E_\text{F}$  is narrower in
Cu-45124(Cl) than in Cu-2252(Cl). The relative proportion of the
Cl-$p$ and O-$p$ contribution to the bands at $E_\text{F}$  is also
smaller in the case of Cu-45124(Cl). The O-$p$ and Cl-$p$ dominated
bands, instead of being separated, overlap to a larger extent in the
case of Cu-45124(Cl). Understanding and quantifying these differences
requires the analysis of the bandstructure in terms of a microscopic
model.


{\it Downfolding and effective model.-} A powerful technique to
construct a low-energy, tight-binding (TB) Hamiltonian starting from a
complex LDA bandstructure is achieved via the {\it N}MTO-downfolding
technique. It does so by constructing energy dependent, effective
orbitals by integrating out irrelevant degrees of freedom - a method
called {\it downfolding}.  The accuracy of such a procedure can be
tuned by the choice of {\it N}, the number of energy points used in
the {\it N}MTO calculation. For an isolated set of bands, as is the
case in the present study, these effective orbitals define the Wannier
functions corresponding to the Hamiltonian in the downfolded
representation. The real space representation of the downfolded
Hamiltonian in the Wannier function basis gives the information of the
effective hopping matrix elements.

\begin{figure}
\includegraphics[width=8cm,keepaspectratio]{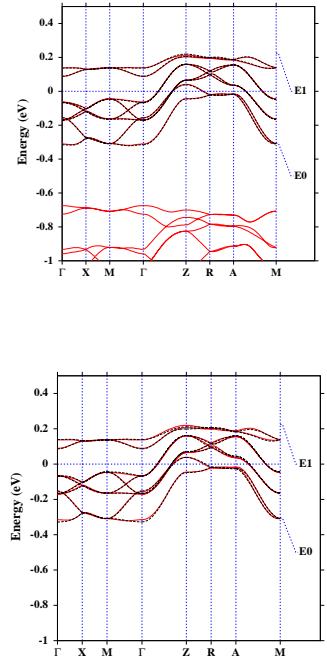}
\caption{
(Color online) Top panel: Bands obtained with massively downfolded Cu
$d_{x^2-y^2}$ basis (in dotted lines) compared to full LDA band
structure (in solid lines).  The {\it N}MTO energy points $E_n$ spanning the
region of interest are shown on the right-hand side.  Bottom panel: The
tight-binding bands obtained with the hopping interactions shown in
Table.~\ref{table} (in dotted lines) compared with downfolded bands
(in solid lines). }
\label{downfold}
\end{figure}

For the present study, we construct the massively downfolded
Hamiltonian by keeping only the Cu-$d_{x^2-y^2}$ degrees of freedom
active and integrating out all the rest. The computed, downfolded
bands are shown in the top panel of Fig.~\ref{downfold} with dotted
lines. With the choice of two energy points, $E_0$ and $E_1$, the
downfolded bands are indistinguishable from the Cu-$d_{x^2-y^2}$
dominated bands of the full LDA calculation shown in solid lines in
the top panel of Fig.~\ref{downfold}.

\begin{figure}
\includegraphics[width=9cm,keepaspectratio]{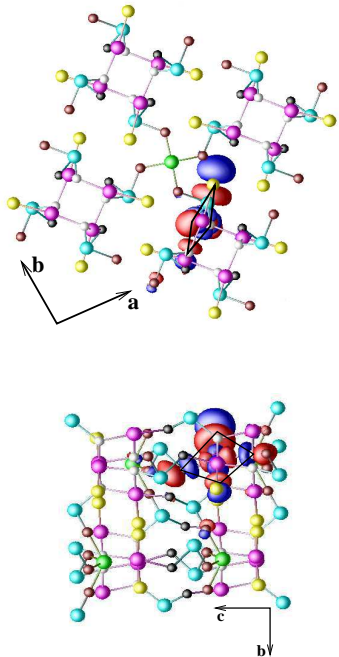}
\caption{(Color online) Effective orbital corresponding to massively downfolded
Cu-$d_{x^2-y^2}$ calculation viewed in two different planes. Plotted
are the orbital shapes (constant-amplitude surfaces) with lobes of
opposite signs colored as red and blue. The $d_{x^2-y^2}$ orbital is
defined with the choice of the local coordinate system with the $y$ axis
pointing along Cu-O(3) and the $x$ axis pointing along the Cu-Cl bond within
the square plane. }
\label{wannier}
\end{figure}

The corresponding Wannier function is plotted in
Fig.~\ref{wannier}. Two different views of the same orbital are
shown. The central part has the $3d_{x^2-y^2}$ symmetry with the
choice of the local coordinate system as stated above, while the tails
are shaped according to Cl-$p_x$ and O-$p_x/p_y$ symmetry
demonstrating the hybridization effects. The strong $pd\sigma$ antibonds
are evident in the plot with Cu hybridization being stronger with Cl
than with O, a fact also evident in the density of states plot, shown
in Fig.~\ref{dos}.

The real space representation of the downfolded Hamiltonian in the
Wannier function basis, $ H_{TB} = - \sum_{ij} t_{ij} ( \hat{c}^{\dagger}_{i}
\hat{c}_{j} + h.c.)$ provides the information of the effective hopping
interaction $t_{ij}$, between the Cu$^{2+}$ ions at sites $i$ and
$j$. The various dominant hopping interactions are tabulated in
Table~\ref{table}. The notation for the hoppings are shown in
Figs.~\ref{struc1} and \ref{struc2}. While the hoppings $t_{1}$,
$t_2$, $t_x$, $t_a$ and $t_{r}$ are in-plane hoppings in the plane
defined by the Cu tetrahedra, $t_z$ and $t_c$ are out-of-plane
hoppings.  For the sake of consistency, we adopt for Cu-45124(Cl) the
same hopping notation introduced earlier for Cu-2252(Cl) in Ref.\
\onlinecite{Valenti03}.

\begin{table}
\begin{tabular}{|l|c|c|c|c|}
\hline
 &   \multicolumn{2}{c} {Cu-45124(Cl)} \vline & \multicolumn{2}{c} {Cu-2252(Cl)} \vline \\
\hline
&{Bond length}&{Interaction }& {Bond length}& {Interaction} \\ \hline
 t$_1$           &  3.147     &    76              &   3.229      &    98 \\ \hline
t$_2$            &  3.523     &     4              &   3.591      &     0 \\ \hline
t$_x$      &  5.539     &    12              &   4.163      &   -10 \\ \hline
t$_a$     &  6.180     &    15              &   6.021      &   -29 \\ \hline
t$_d$  &  7.834     &     7              &   8.033      &   -80 \\ \hline
t$_r$   &  8.251     &    18              &   9.048      &   -48 \\ \hline
t$_z$   &  5.063     &    24              &   5.015      &    12 \\ \hline
t$_c$    &  6.332     &   -48              &   6.320      &   -45 \\ \hline
\end{tabular}
\caption{Cu-Cu hopping  parameters corresponding to the massively downfolded Cu-$d_{x^2-y^2}$
Hamiltonian. The bond lengths are in {{\AA}} and the hopping interaction strengths
are in meV corresponding to hoppings shown in Figs.~\ref{struc1} and
~\ref{struc2}. The values for Cu-2252(Cl) have been reproduced from
Ref.~\onlinecite{Valenti03}.}
\label{table}
\end{table}

For comparison, in Table~\ref{table} we reproduce the results for the
Cu-2252(Cl) compound from Ref. \onlinecite{Valenti03}. The bond
lengths corresponding to each hopping element have been also
tabulated.  For the dominant, intratetrahedral nearest neighbor
interaction, $t_1$, we observe that while the bond length is decreased
by only $2.5\%$ for Cu-45124(Cl) compared to Cu-2252(Cl), $t_1$ is
reduced by as much as $22\%$ due to a smaller superexchange path angle
in Cu-45124(Cl) ($\angle$ Cu-O(2)-Cu $=105.7${{\textdegree}}{} for Cu-45124(Cl) and $\angle$
Cu-O(1)-Cu$= 109.8${{\textdegree}}{} for Cu-2252(Cl)). 
The intratetrahedral hopping $t_2$ which was weak
for Cu-2252(Cl) -a fact also supported by neutron
diffraction~\cite{Zaharko04}-, remains weak for Cu-45124(Cl). The
in-plane intertetrahedral hopping $t_x$, remains in magnitude similar
to its analog in Cu-2252(Cl) while other in-plane intertetrahedral
hoppings like $t_a$ and $t_r$ get suppressed. The out-of-plane,
intertetrahedral hopping, $t_c$ remains more or less the same as in
Cu-2252(Cl), while the $t_z$ hopping increases by a factor of two. The
most remarkable change is observed for the diagonal hopping,
$t_d$, which is reduced to 7~meV in Cu-45124(Cl) compared to a value
of 80~meV in the Cu-2252(Cl) compound. This reduction however is not
caused by the elongation of the bond lengths due to the
insertion of the Te(1)O$_4$ group in Cu-45124(Cl), as was suggested in
Ref. \onlinecite{peter}.  We reveal the origin of this marked
difference in the following in terms of a detailed analysis of the
involved hopping paths.

The tight-binding (TB)  bands, constructed out of the hopping parameters
tabulated in Table~\ref{table} are shown on the bottom  panel of
Fig.~\ref{downfold} in comparison to downfolded bands. The TB bands
compare satisfactorily with the downfolded bands.  Omission of
long-ranged interactions such as $t_c$ and $t_r$ deteriorates the
agreement of the TB bands with the downfolded bands, proving the
essential need for inclusion of long-ranged interactions in the
correct description of this compound.


\begin{figure}
\includegraphics[width=10cm,keepaspectratio]{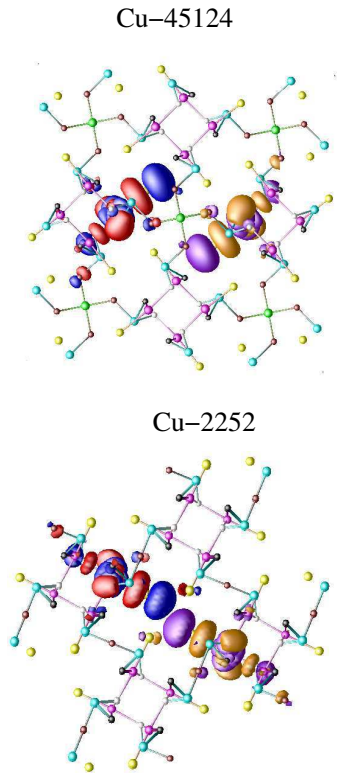}
\caption{(Color online) Overlap between Cu-$d_{x^2-y^2}$ downfolded {\it N}MTOs,
placed at two Cu sites situated at in-plane intertetrahedral,
diagonal positions. Opposite signed lobes of the orbitals are colored
as blue or magenta and red or orange.  }
\label{ovl}
\end{figure}
{\it Interaction pathways.-} It was pointed out in Ref.~
\onlinecite{Valenti03} that Cl-$p$ degrees of freedom play a crucial
role in the renormalization process of the effective Cu-Cu
hopping. Keeping this fact in mind, we carried out downfolding
calculations where the Cl-$p$ degrees of freedom have been kept active
in addition to Cu-$d_{x^2-y^2}$, so as to define a basis consisting of
Cu-$d_{x^2-y^2}$ and Cl-$p$. The Cu-Cu hopping interactions extracted
out of such calculation are tabulated in Table.~\ref{table1}. For
comparison, we show the results for Cu-2252(Cl) reproduced from
Ref.~\onlinecite{Valenti03}. The crucial role of hopping paths
involving Cl-$p$ is evident by comparing the hopping interactions
between the massively downfolded Cu-$d_{x^2-y^2}$-only calculation and
the Cu-$d_{x^2-y^2}$ + Cl-$p$ calculation. The former includes the
renormalization due to Cl-$p$'s while the latter does not. While the
pattern of renormalization remains essentially the same for the
intratetrahedral hopping $t_2$ and the intertetrahedral out-of-plane
hopping $t_c$, it is quite different for intertetrahedral hoppings
like $t_x$, $t_a$, $t_d$, $t_r$ and $t_z$ which involve pathways via
Cl atoms belonging to two different Cu$_4$ tetrahedra. The most
significant change happens for the in-plane intertetrahedral diagonal
hopping, $t_d$. The bare hopping strength of $t_d$ in absence of the
renormalization effect of Cl-$p$ is more or less the same between the
two compounds (8~meV for Cu-2252(Cl) and 9~meV for
Cu-45124(Cl)). However, while a large renormalization is observed for
Cu-2252(Cl) when integrating out the Cl-$p$ degrees of freedom, such
renormalization is practically absent in Cu-45124(Cl). This difference
is caused -as pointed out previously- by the different alignment of
the Cu-Cl bonds belonging to neighboring Cu$_4$ tetrahedra which are
parallel to each other in Cu-45124(Cl) while in Cu-2252(Cl) they point
to each other.  {\it This makes the intertetrahedral Cl-$p$ - Cl-$p$
bonding in Cu-45124(Cl) of $pp\pi$ type as opposed to the Cu-2252(Cl)
case, where the Cl-$p$ - Cl-$p$ bonding was of $pp\sigma$ type}. This is
nicely demonstrated in the Wannier function plot (see Fig.~\ref{ovl}),
where the effective Cu-$d_{x^2-y^2}$ like Wannier orbitals are placed
at the Cu sites at in-plane diagonal positions. The overlap between
the two orbitals provides a rough estimate of the strength of the
hopping matrix elements.  In the case of Cu-2252(Cl), the Cl-$p$ tails
from two Cu sites belonging to two different Cu$_4$ tetrahedra overlap
to a large extend due to direct alignment providing a Cl-$p$-Cl-$p$
$pp\sigma$ bonding which mediates the Cu-Cu bonding between different
Cu$_4$ tetrahedra.  For Cu-45124(Cl), in contrast, the overlap of the
Cl-$p$ tails from different Cu sites belonging to two different Cu$_4$
tetrahedra is practically negligible due to misalignment of the Cl
orbitals.

\begin{table}
\begin{tabular}{|r|r|r|r|r|r|}
\hline
 &   \multicolumn{2}{c} {Cu-45124(Cl)} \vline & \multicolumn{2}{c} {Cu-2252(Cl)} \vline \\
\hline
        & Cu      &   Cu+Cl  & Cu        & Cu+Cl \\
\hline
 t$_1$    &   76       &    82    &     98        &    181      \\
\hline
t$_2$     &    4       &  -117    &      0        &   -132     \\
\hline
t$_x$    &   12       &    42    &    -10        &    -14     \\
\hline
t$_a$    &   15       &   -39    &    -29        &      8     \\
\hline
t$_d$    &    7       &     9    &    -80        &      8     \\
\hline
t$_r$    &   18       &   -11    &    -48        &    -72   \\
\hline
t$_z$    &   24       &    27    &     12        &     33   \\
\hline
t$_c$    &  -48       &   -15    &    -45        &    -19  \\
\hline
\end{tabular}
\caption{TB parameters in meV corresponding to two sets of calculations. Set-1:
Massively downfolded Cu-$d_{x^2-y^2}$, Set-2: minimal set consisting
of Cl-$p$ and Cu-$d_{x^2-y^2}$ degrees of freedom (Cu+Cl downfolding).
The numbers for Cu-2252(Cl) have been reproduced from
Ref.~\onlinecite{Valenti03}}
\label{table1}
\end{table}

The nature of the discussed interaction paths plays a crucial role in
the magnetic properties of this material.  Starting from the hopping
parameters, $t$'s, the exchange integrals, $J$'s, for
antiferromagnetic superexchange paths may be estimated by making use
of the expression $J \approx 4t^2/U$. While this is a valid approach for
cases like the $t_1$ and $t_2$ interaction paths, in general for more complicated
paths this expression is not anymore precise and one has to use more
involved estimations of the exchange coupling constants. Nevertheless,
already the knowledge of the hopping parameters gives us the clue
about the important interaction paths.   The
drastic reduction of the in-plane $t_d$ and to a lesser extent of
$t_a$ and the longer-ranged $t_r$, in Cu-45124(Cl) compared to
Cu-2252(Cl), indicate an overall weakening of the intertetrahedral
coupling in the new compound with respect to Cu-2252(Cl) and therefore
if the system orders at low temperatures, the ordering should occur at
a lower T$_N$ than in Cu-2252(Cl), as observed
experimentally~\cite{peter}.
 The spin ordering patterns
will be also strongly influenced by the change of interaction paths,
especially by the near absence of the $t_d$ and reduction of the $t_a$
path (we refer to the discussion in Ref. \onlinecite{neutron}) which
places the system in the limit of weakly coupled tetrahedra.  Also the
reduction by 22\% of the $t_1$ value implies a smaller intratetrahedron
exchange coupling $J_1$ than in Cu-2252(Cl). By considering $J \approx
4t^2/U$,
with U=4eV we obtain as exchange coupling constants $J_1 \approx 5.8 meV =
67 K$ and  $J_2 \approx 0.02 meV = 0.2 K$ in comparison with the values
$J_1 = 2.84 meV =
32.9 K$ and  $J_2 = 1.58 meV = 18.4 K$ obtained by fitting the susceptibility
 of a model of independent tetrahedra to the experimental data\cite{peter}.
The ratio of $J_2/ J_1$ is largely overestimated in the fitting, presumably 
because of the neglect of the inter-tetrahedral interactions.


{\it Br System.-} In an attempt to predict the properties of the not
yet synthesized Cu$_4$Te$_5$O$_{12}$Br$_4$ (Cu-45124(Br)) and
motivated by the discussed proximity to a quantum critical behavior of
the much studied Cu$_{2}$Te$_{2}$O$_{5}$Br$_{2}$ (Cu-2252(Br)), we
have investigated the electronic and magnetic properties of the {\it
ab initio} relaxed structure Cu-45124(Br) obtained from first
principles calculations. In order to obtain a theoretical prediction
of the Cu-45124(Br) crystal structure, we substituted Cl by Br in the
original Cu-45124(Cl) structure and we relaxed the volume and internal
coordinates performing Car-Parrinello {\it ab initio} molecular
dynamics (AIMD) calculations~\cite{AIMD} with a projector augmented
wave (PAW) basis set~\cite{PAW}.  This procedure has proven to be very
suitable for predicting reliable new crystal
structures~\cite{Salguero06}.  For Cu-45124(Br) we assume the same
tetragonal space group $P4/n$ (No. 85) as for Cu-45124(Cl). Of the
seven atoms in the primitive cell, only Te(1) is in Wyckoff position
$2c$, while all others (Te(2), Cu, Br, O(1), O(2), O(3)) are in
position $8g$. We thus have 19 degrees of freedom, but as the AIMD
relaxation is done in the conventional cell, we need 131 constraints
for the 50 atoms in order to preserve the symmetry. We verify
convergence of our structure relaxation not only with the help of the
forces but we check that each of the 19 independent coordinates has
converged.
This is especially important in this structure as we find that the relaxation happens in two steps: First, the Br atoms introduced into the Cl positions rearrange, increasing their bond distance to the Cu atoms which are the nearest neighbors. This Cu-Br repulsion makes an adjustment of the O(1) positions next to Br and the O(2) and O(3) positions next to Cu necessary. But as soon as the Br atom has found a relatively favorable position, some changes to Cu and O(1)-O(3) coordinates are actually reversed. Thus, the structure immediately following the first fast rearrangement would have produced quite different interactions strengths than the final relaxed structure given in Table~\ref{Brcoordinates}.
 As would be expected from the different radii of Br and Cl
atoms, we find the largest adjustments in the Br atom positions which
change by 0.16~{{\AA}}{} during the relaxation. The other changes are
0.05~{{\AA}}{} for Te(1), 0.04~{{\AA}}{} for Te(2), 0.04~{{\AA}}{} for Cu,
0.10~{{\AA}}{} for O(1), 0.04~{{\AA}}{} for O(2) and 0.07~{{\AA}}{} for O(3).

\begin{table}
\begin{tabular}{|c|c|c|c|}
\hline
 & $x$ & $y$ & $z$  \\
\hline
Te(1)  & $0.25         $ & $0.25          $ & $0.37660383$\\
Te(2)  & $0.67370804   $ & $0.018727271   $ & $0.87086473$\\
Cu     & $0.75587933   $ & $0.40510264    $ & $0.34817424$\\
Br     & $0.89355065   $ & $0.56680922    $ & $0.32307512$\\
O(1)   & $0.29531785   $ & $0.40609792    $ & $0.23273845$\\
O(2)   & $0.28384154   $ & $0.87238649    $ & $0.36101169$\\
O(3)   & $0.291597     $ & $0.58080843    $ & $0.93893775$\\
\hline
\end{tabular}
\caption{Fractional coordinates obtained by AIMD
 of the relaxed Cu-45124(Br).}
\label{Brcoordinates}
\end{table}

While the volume of the new structure shows a negligible change with
respect to the volume of Cu-45124(Cl), appreciable changes in bond
lengths and angles are observed.  The Cu-O(2) distance which
alternates between 1.94~{{\AA}}{} and 2.01~{{\AA}}{} in Cu-45124(Cl), becomes
1.91~{{\AA}}{} and 2.01~{{\AA}}{} in Cu-45124(Br). The Cu-O(3) distance
is slightly smaller at 1.90~{{\AA}} (from 1.91~{{\AA}} in Cu-45124(Cl)) . The Cu-O(2)-Cu angle changes from
105.7{{\textdegree}}{} in Cu-45124(Cl) to 107.2{{\textdegree}}{} in Cu-45124(Br), while the O-Cu-O angle stays nearly constant at 87.0{{\textdegree}}{} (87.1{{\textdegree}}{} in Cu-45124(Cl)). The Cu-Cl
distance is 2.24~{{\AA}}, the Cu-Br distance 2.42~{{\AA}}. Finally, the
Cl-O(1) distances alternate between 3.27~{{\AA}}{} and 3.40~{{\AA}}{} while
the Br-O(1) distances are 3.22~{{\AA}}{} and 3.55~{{\AA}}.  In
Table~\ref{Brcoordinates} we present the relaxed coordinates of
Cu-45124(Br).

We performed {\it N}MTO-downfolding for this system and in
Tables~\ref{Brhopdis} and \ref{Brhoppings} we present the bond
distance and hopping values together with those of (Cu-2252(Br)). Both
a Cu-$d_{x^2-y^2}$ and a Cu-$d_{x^2-y^2}+$Br-$p$ downfolding were
performed. 

\begin{table}
\begin{tabular}{|l|c|c|c|c|}
\hline
 &   \multicolumn{2}{c} {Cu-45124(Br)} \vline & \multicolumn{2}{c} {Cu-2252(Br)} \vline \\
\hline
&{Bond length}&{Interaction }& {Bond length}& {Interaction} \\ \hline
 t$_1$           &  3.147     &    75             &   3.195      &    80 \\ \hline
t$_2$            &  3.522    &     0             &   3.543      &     4 \\ \hline
t$_x$      &  5.535   &    23              &   4.385     &   -16 \\ \hline
t$_a$     &  6.248   &    20              &   6.289      &   -30 \\ \hline
t$_d$  &  7.829     &     3              &   8.439      &   -73 \\ \hline
t$_r$   &  8.251    &    -29              &   9.130      &   -35 \\ \hline
t$_z$   &  5.064    &    19              &   5.059      &    11 \\ \hline
t$_c$    & 6.332    &   -39              &   6.378      &   -48 \\ \hline
\end{tabular}
\caption{Cu-Cu hopping parameters  corresponding to the massively downfolded Cu-$d_{x^2-y^2}$
Hamiltonian. The bond lengths are in {{\AA}} and the hopping interaction strengths
in meV corresponding to hoppings shown in Figs.~\ref{struc1} and
\ref{struc2}. The numbers for Cu-2252(Br) has been reproduced from
Ref.~\protect\onlinecite{Valenti03}.}
\label{Brhopdis}
\end{table}

\begin{table}
\begin{tabular}{|r|r|r|r|r|r|}
\hline
 &   \multicolumn{2}{c} {Cu-45124(Br)} \vline & \multicolumn{2}{c} {Cu-2252(Br)} \vline \\
\hline
        & Cu      &   Cu+Br  & Cu        & Cu+Br \\
\hline
 t$_1$   &   75      &    106    &     80       &    155      \\
\hline
t$_2$    &    0       &  -60   &      4        &   -156     \\
\hline
t$_x$    &   23       &   -15    &    -16        &    -10     \\
\hline
t$_a$    &   20       &   -16    &    -30        &      5     \\
\hline
t$_d$    &    3       &    11    &    -73        &      8     \\
\hline
t$_r$    &   -29       &   -16    &    -35        &    -62   \\
\hline
t$_z$    &   19       &    68    &     11        &     34   \\
\hline
t$_c$    &  -39       &   -53    &    -48        &    -26  \\
\hline
\end{tabular}
\caption{Cu-Cu hopping parameters in meV corresponding to two sets of calculations. Set-1:
Massively downfolded Cu-$d_{x^2-y^2}$, Set-2: minimal set consisting
of Br-$p$ and Cu-$d_{x^2-y^2}$ degrees of freedom (Cu+Br downfolding).
The numbers for Cu-2252(Br) have been reproduced from
Ref.~\onlinecite{Valenti03}.}
\label{Brhoppings}
\end{table}

Cu-45124(Br) shows the same trend as Cu-45124(Cl) regarding the
intertetrahedral hopping $t_d$, namely the near absence of Cu-Cu
interaction along this path.  The rest of in-plane intertetrahedral hopping
paths in Cu-45124(Br) are a bit larger than in Cu-45124(Cl) but,
except for $t_x$, they are smaller than in Cu-2252(Br).   From the
knowledge of the previous systems, a phase transition to an ordered
state is also to be expected for this system at low temperatures.

An important issue to be mentioned at this point is the value of the
intratetrahedron ratio $t_2/t_1$ in all the compounds discussed here.
Large values of this ratio can be related to an enhancement of
intratetrahedron frustration, what has been already discussed for
Cu-2252(Br)\cite{Valenti03}. Cu-2252(Br) is found to have a small 
but nonzero $t_2$ in comparison to its value for Cu-2252(Cl), where the 
$t_2$ hopping path
is basically zero, mainly due to the Cl renormalization.  The new set
of systems, i.e the synthesized Cu-45124(Cl) and the {\it ab initio}
computer designed Cu-45124(Br) seem to behave in the opposite way.
While Cu-45124(Cl) has a small but nonzero $t_2$, Cu-45124(Br) relaxes
into a structure where the $t_2$ path is completely renormalized to
zero by the hybridization with the Br ions  (see Table \ref{Brhoppings})
in the square planar
configuration.  Though we found an interesting transient structure for
Cu-45124(Br) with a moderate intratetrahedron $t_2/t_1$, this doesn't
seem to be the energetically favored structure within the AIMD
approach.

Finally, we would like to note that the value of the {\it ab initio}
calculated hopping parameters is very susceptible to small changes of
distances and angles between the atoms.  Our AIMD calculations were
performed, as mentioned previously, within the GGA approximation.
Consideration of other exchange correlation potentials may change
slightly the relaxed structure, which could be important especially
for the intratetrahedron hopping paths, where changes of 0.02 to 0.03
{\AA} in the distance between Cu and O(2) and of 2.9{\textdegree} in the
Cu-O(2)-Cu angle are decisive for the variation of the hopping
parameters.

{\it Summary.-} To conclude, we have made a comparative study between
the spin tetrahedron system Cu$_{2}$Te$_{2}$O$_{5}$Cl$_{2}$ and a
newly synthesized compound Cu$_4$Te$_5$O$_{12}$Cl$_4$ belonging to the
same family, in terms of the microscopic analysis of the electronic
structure. Our study shows that although the basic nature of the
interactions remains the same, there is a drastic reduction of the
in-plane intertetrahedral diagonal interaction in comparison to the
case of Cu$_{2}$Te$_{2}$O$_{5}$Cl$_{2}$ where this diagonal
interaction was estimated to be nearly as strong as the Cu$_4$
intratetrahedral nearest neighbor interaction $t_1$. We show that the
origin of this reduction is due to subtle changes in the crystal
structure of Cu$_4$Te$_5$O$_{12}$Cl$_4$ which causes Cu-Cl bonds
belonging to different Cu$_4$ tetrahedra to align in parallel in the
new compound rather than pointing towards each other as was the case
in Cu$_{2}$Te$_{2}$O$_{5}$Cl$_{2}$.  This reduction of the in-plane
diagonal hopping in turn increases the importance of the out-of-plane
hopping to the extent that some intertetrahedral hoppings along [001] (t$_c$) are even 
about three times stronger than those within the plane.


In absence of the, yet to be synthesized, Cu$_4$Te$_5$O$_{12}$Br$_4$
and motivated by the more anomalous properties observed in the Br
analog to Cu$_{2}$Te$_{2}$O$_{5}$Cl$_{2}$, we have theoretically
derived the hypothetical crystal structure of
Cu$_4$Te$_5$O$_{12}$Br$_4$ by performing a geometry relaxation in the
framework of {\it ab initio} molecular dynamics.  We have analyzed the
electronic properties of this new system within the {\it N}MTO
downfolding procedure. We observe that, while the overall electronic
and magnetic behavior seems to be similar to its Cl sister compound,
this computer designed Br system shows (except for $t_d$)  a stronger in-plane
intertetrahedron interaction than the Cl system  -an effect that was also observed in the
 comparison between Cu$_{2}$Te$_{2}$O$_{5}$Br$_{2}$ and Cu$_{2}$Te$_{2}$O$_{5}$Cl$_{2}$-. 
We, however  don't observe any noticeable effect on the intratetrahedron frustration in
the final relaxed  Cu$_4$Te$_5$O$_{12}$Br$_4$ structure.


{\it Acknowledgments.-} We would like to thank P. Lemmens and
M. Johnsson for very fruitful discussions. R.V. thanks the German
Science Foundation (DFG) for financial support. B.R. and T.S.D thanks
the MPG-India partnergroup program for the
collaboration. H.O.J. gratefully acknowledges support from the DFG
through the Emmy Noether program. We gratefully acknowledge support
by the Frankfurt Center for Scientific Computing.


\begin{thebibliography}{99}
\bibitem{comment_data} Please note that we used the atom labelling given in
 Refs. \protect\onlinecite{Johnsson00,peter}.
\end{thebibliography}
\end{document}